\documentclass[preprint,12pt]{elsarticle}
\usepackage{lineno}
\usepackage{mathrsfs}
\usepackage{amsmath}
\usepackage{color}
\usepackage{bbm}
\usepackage{cases}
\usepackage{amsfonts}

\usepackage{amssymb}
\usepackage{booktabs}
\usepackage{graphicx}
\usepackage{cases}
\usepackage{epstopdf}
\usepackage{setspace}
\usepackage{subfigure}
\usepackage{indentfirst} 
\usepackage{hyperref}  
\usepackage{ragged2e}
\usepackage{url} 

\usepackage{algorithm}
\usepackage{algorithmicx}
\usepackage{algpseudocode}
\usepackage{amsmath}
\graphicspath{{}}
\usepackage{lineno}
\usepackage{amsthm}

\newcommand\old[1]{}

\usepackage{graphicx}
\biboptions{numbers,sort&compress}

\usepackage{geometry}
\usepackage{setspace}

	\usepackage{caption}

\modulolinenumbers[5]

\journal{Applied Soft Computing}

\begin{document}
\sloppy

\begin{frontmatter}
	\captionsetup[figure]{labelfont={bf},labelformat={default},labelsep=period,name={Fig.}}
	
	\title{Semi-supervised object detection based on single-stage detector for thighbone fracture localization}
	\begin{abstract}
 The thighbone is the largest bone supporting the lower body. If the thighbone fracture is not treated in time, it will lead to lifelong inability to walk. Correct diagnosis of thighbone disease is very important in orthopedic medicine. Deep learning is promoting the development of fracture detection technology. However, the existing computer aided diagnosis (CAD) methods baesd on deep learning rely on a large number of manually labeled data, and labeling these data costs a lot of time and energy. Therefore, we develop a object detection method with limited labeled image quantity and apply it to the thighbone fracture localization. In this work, we build a semi-supervised object detection(SSOD) framework based on single-stage detector, which including three modules: adaptive difficult sample oriented (ADSO) module, Fusion Box and deformable expand encoder (Dex encoder). ADSO module takes the classification score as the label reliability evaluation criterion by weighting, Fusion Box is designed to merge similar pseudo boxes into a reliable box for box regression and Dex encoder is proposed to enhance the adaptability of image augmentation. The experiment is conducted on the thighbone fracture dataset, which includes 3484 training thigh fracture images and 358 testing thigh fracture images. The experimental results show that the proposed method achieves the state-of the-art AP in thighbone fracture detection at different labeled data rates, \textit{i.e.} 1\%, 5\% and 10\%. Besides, we use full data to achieve knowledge distillation, our method achieves 86.2\% AP50 and 52.6\% AP75.
	\end{abstract}
	
	\author[mymainaddress]{Jinman Wei}
	\ead{2021234147@tju.edu.cn}
	\author[mysecondaddress]{Jinkun Yao}
	\ead{yjk1213@163.com}
	\author[mymainaddress]{Guoshan Zhang\corref{mycorrespondingauthor}}
	\ead{zhanggs@tju.edu.cn}
	\author[mymainaddress]{Bin Guan}
	\ead{guanbin@tju.edu.cn}
	\author[mymainaddress]{Yueming Zhang}
	\ead{seife@tju.edu.cn}
		\author[mymainaddress]{Shaoquan Wang}
	\ead{sqwang@tju.edu.cn}
	
	\cortext[mycorrespondingauthor]{Corresponding author}
	\address[mymainaddress]{School of Electrical and Information Engineering, Tianjin University, Tianjin, 300072, China.}
	\address[mysecondaddress]{Department of Radiology,Linyi People's Hosptial,276000 Linyi,China.}
\end{frontmatter}

\section{Introduction}
The thighbone is located below the pelvis. Thighbone and acetabulum constitute the hip joint and play a role in supporting the whole body. Various activities of the human body depend on the thighbone, so it is one of the most vulnerable part. The diagnosis of ordinary fracture and comminuted fracture is a significant part of surgical diagnosis\cite{2020Assessment}. However, compared with the huge number of patients, there is a lack of excellent surgeons. Therefore, surgeons urgently need an assistant to relieve their work pressure. In order to solve this problem, many computer-aided detection and diagnosis  methods\cite{georgalis2022crushed} have been proposed. In recent years, substantial progress has been made in developing deep learning-based CAD systems to fracture diagnosis. Guan \textit{et al.} proposed a convolutional neural network for thighbone fracture detection that can balance the information of each feature map in ResNeXt's feature pyramid.\cite{0Automatic}. Firat \textit{et al.} designed an integrated object detection model for wrist X-ray image fracture detection\cite{hardalacc2022fracture}. At present, the state-of-the-art  fracture detection methods are usually developed based on large-scale expert annotations such as 5134 labeled CT images for spinal fracture detection\cite{sha2020detection}, 7356 wrist radiographic images\cite{thian2019convolutional}, 9040 labeled  hand, wrist, knee, ankle, foot and ankle radiographs for multiple fracture detection\cite{wu2021feature}.\par
Compared with the above-mentioned methods, semi-supervised learning (SSL) uses both labeled data and unlabeled data when
training the model, and uses unlabeled data to assist in optimizing the model, 
so as to save training cost. The state-of-the-art semi-supervised methods are the pseudo-label based approaches\cite{Lee2013}. Specifically, the model is trained on labeled data, 
and then the trained model is used to predict the pseudo labels on unlabeled images. Teacher-student model\cite{2015Distilling} is a 
common method to generate pseudo labels in semi-supervised learning in which key idea is to train two independent models, namely teacher model and student model. 
The teacher model is trained on the labeled images to label the unlabeled images and then mix these pseudo labeled images with the labeled images to train the student model.\par
Most research on SSOD has focused on the two-stage detectors\cite{cai2018, wang2019region, zhu2019empirical, cao2019gcnet}. But basing on the single-stage detectors (such as FCOS\cite{tian2019fcos}, YOLOF\cite{chen2021you}, RestinaNet\cite{lin2017focal}) has more attractive and practical, because they can be easily deployed on devices with limited resources, eliminate cumbersome preprocessing and post-processing except for NMS\cite{Redmon2016}. The main difference between the single-stage detectors and the two-stage detectors is that Region proposal network (RPN)\cite{2016Faster} of the two-stage detector can filter most of the background samples, and in the next stage, the remaining candidate boxes are further predicted the detailed categories. The single-stage detectors make dense prediction for all areas of the image at one time, as long as few bounding boxes can be predicted as positive samples. Because in the single-stage detectors, the generation and judgment of the proposal are integrated, this lead to that the detection speed is faster but the classification score of one-stage detectors is lower than two-stage detectors. And directly sending pseudo labels with low classification score into student model will bring a lot of noise and affect the training accuracy of the model. Therefore, how to deal with a large number of low-quality pseudo labels in dense prediction is still an important problem.\par
Regression branch is another component of object detection task. The regression quality of pseudo box is another important factor that determines the performance of semi supervised target detection model.  Xu \textit{et al.}\cite{Xu2021} find that the accuracy of the regression is related to the uncertainty calculated by the BoxJitter module, but the BoxJitter module relies on Regions with CNN features(RCNN) to process the proposal, so it is not applicable to the single-stage detector. To address this issue, we propose the Fusion Box module in the regression branch for SSOD based on single-stage detector.\par
In summary, we develop a semi-supervised framework based on the single-stage detector for the thighbone fracture detection. In this framework, the adaptive difficult sample oriented(ADSO) module and the Fusion Box module are developed to reduce the impact of inaccurate pseudo label prediction. In addition, The Single-in-Single-out (SISO) encoder called Dex encoder is proposed to improve the adaptability of the augmented input images. The main contributions of this paper can be summarized as follows:\par
1. We developed the semi-supervised object detection framework based on single-stage detector for thighbone fracture detection with limited annotations. Compared with previous work, it has fewer parameters and faster detection speed.\par
2. The adaptive difficult sample oriented (ADSO) module is proposed to take the classification score of teacher model as the criterion of pseudo labels reliability.\par
3. The Fusion Box module is proposed to reduce the impact of multiple pseudo boxes regression in the same position on model performance.\par
4. We design a Single-in-Single-out encoder named Deformable expand encoder (Dex encoder) for enhancing the learning ability of of deformed features.\par
5. The experimental results show that compared with supervised and semi-superviesed methods, our method is better than other methods in thighbone fracture detection.
\section{Related work}
\subsection{Deep learning for medical detection} 
CAD has been extensively studied in the past decade\cite{hesamian2019deep,fujita2020ai}, and CAD system based on deep learning has been developed to diagnose a wide range of Pathology such as detection of covid-19\cite{Karthik2021,gupta2022covid}, mass and calcification features in mammography\cite{2018Simultaneous} and brain tumor diagnosis\cite{ma2021}. In the fracture detection method based on deep learning\cite{2020A}, FAMO\cite{wu2021feature} constructed the Feature Ambiguity Mitigate Operator model to mitigate feature ambiguity in bone fracture detection on radiographs of various body parts. Due to the requirements of medical professional knowledge, the labor cost of large-scale annotations is expensive which hinders the development of CAD solutions based on deep learning. Computer aided detection using SSL method is an emerging task in recent years, such as Yirui Wang \textit{et al.} proposed the adaptive asymmetric label sharpening (AALS) algorithm using the teacher-student model paradigm, which solves the label imbalance problem unique to the medical field\cite{Wang2021}.\par
\subsection{object detection}
\begin{figure}
	\centering 
	\includegraphics[scale=0.4]{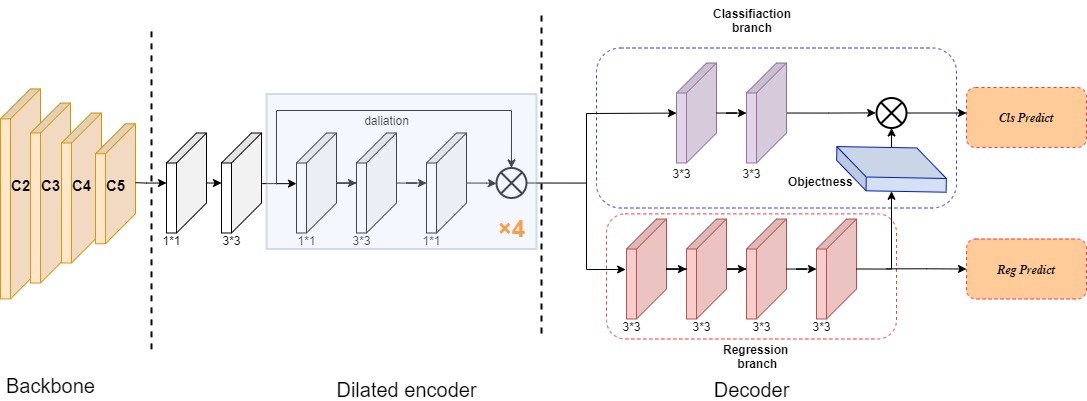}
	\caption{The structure of YOLOF.}
	\label{figure20}
\end{figure}

Object detection is one of the core tasks in computer vision. At present, the object detector based on CNN can be divided into single-stage and two-stage detectors. FasterRcnn\cite{2016Faster} is the representative two-stage detector, which uses RPN network for proposal extraction and RCNN head for regional prediction and extraction of objects. The single-stage detector only uses the features extracted by the feature extraction network for regression and classification. For example, SSD\cite{Liu2016} uses the feature pyramid method to complete target regression and classification on different scale features at the same time. Chen \textit{et al.} developed the YOLOF that only uses C5 feature for detection as shown in Figure \ref{figure20} in which the complex Multiple-in-Multiple-out encoder is replaced by the simple Single-in-Single-out encoder, YOLOF containing two key components: dilated encoder and uniform decoder.
\subsection{Semi-supervised learning in object detection}
SSL method plays a leading role in image classification\cite{2016Temporal,2017Mean,2019ReMixMatch,Sohn2020,pourpanah2021semisupervised}. Because the object detector has complex architecture design 
and multi task learning (classification and regression), 
it is not a simple work to transfer the SSL 
method to the object detection task. 
The current SSOD method mainly has two directions: 
Consistency Regularization\cite{jeong2019consistency} 
and Pesudo Label\cite{Lee2013}. The former uses two deep convolution 
neural networks to learn the consistency between different 
data augmentation\cite{zoph2020learning} 
(horizontal flip, different contrast, brightness, etc.) 
of the same unlabeled image, and make the image prediction to small disturbance the same. The latter uses the 
pre-training model learned on labeled data to infer the 
unlabeled data. In recent years, semi-supervised object 
detection method has attracted people's attention\cite{yang2021interactive,wang2021data,2022S4OD}. STAC\cite{sohn2020simple} 
first applies pseudo label method to SSOD, it apply 
weak data augmentation to unlabeled data, and uses the 
trained teacher model to generate pseudo labels of 
unlabeled images. Unbiased teacher\cite{2021Unbiased} uses focal loss\cite{lin2017focal} to solve the imbalance 
between positive and negative samples. 
Instant teaching\cite{zhou2021instant} 
trains two models at the same time to 
check and correct pseudo labels for 
each other, so as to effectively suppress 
the accumulation of false predictions. Almost all the above work is based on the two-stage detector, such as FasterRcnn, which is not convenient to develop in the medical field with limited resources. Inspired by the above works, we designed a fast semi-supervised detection model based on the single-stage detector.\par

\section{Methology}
\begin{figure}
	\centering 
	\includegraphics[scale=0.3]{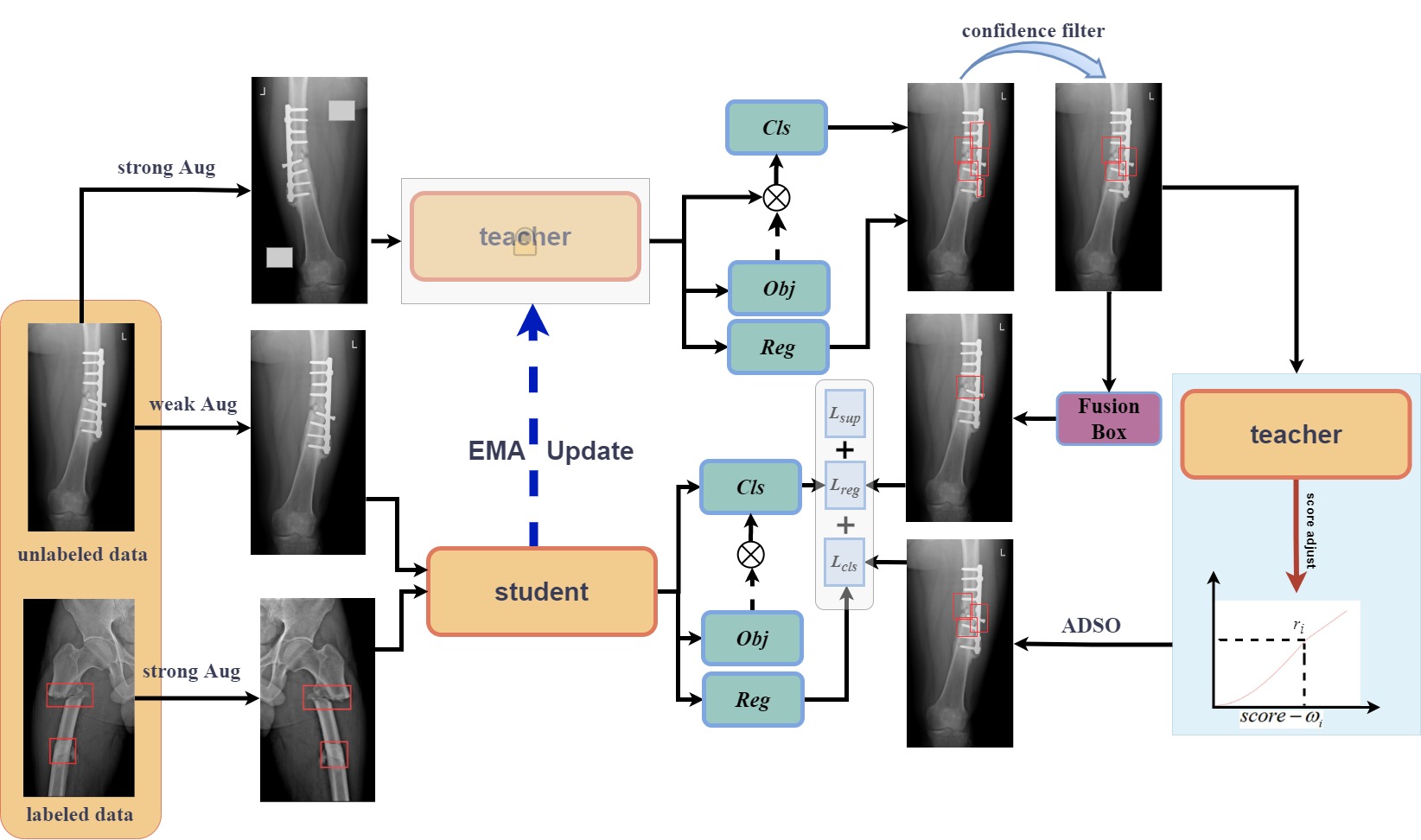}
	\caption{The pipeline of established semi-supervised object detection framework: the labeled images and unlabeled images are sent into the training pipeline in batches. The teacher labels the unlabeled images with pseudo labels as student's ground truth and the teacher does not back propagate. The student model adopts EMA to transfer parameters and update the teacher model. The ADSO of classification (Cls) branch adjusts the confidence of the pseudo labels to evaluate the reliability of the pseudo label. The regression(Reg) branch judges whether to merge the pseudo boxes according to similarity $\xi$. The loss function of classification branch and regression branch adopts
    focal loss and CIOU loss respectively.}
	\label{figure0}
\end{figure}

Our method adopts the teacher-student mutual learning mode in which the student model learns from the detection loss of labeled and unlabeled images. The unlabeled images have two groups of pseudo boxes, which are used for classification branch and regression branch training, respectively. The teacher model is updated by using the student model with exponential moving average (EMA). The pseudo boxes predicted by the teacher model will be filtered by confidence at first, and then the pseudo labels with classification scores higher than the confidence threshold  $\sigma$ will be retained. The remaining pseudo boxes will be sent to the classification branch and regression branch. In this SSOD framework, There are two critical designs: ADSO and Fusion Box. Figure \ref{figure0} shows the description of our SSOD framework.\par
\subsection{Semi-supervised learning framework} 
In each training iteration, unlabeled images and labeled images are extracted according to a certain data sampling ratio. The data are preprocessed by two different preprocessing methods to obtain strong augmented labeled images, weak augmented and strong augmented unlabeled images. The student network is trained with the pseudo boxes generated by teacher model and ground truth boxes in labeled images. The total loss function can be expressed by (\ref{math1}) :\par
\begin{equation}\label{math1}
	L=L_{\text {sup }}+\lambda L_{\text {unsup }}
\end{equation}
where $L_{\text {sup }}$ and $L_{\text {unsup}}$ represent the loss function of labeled images and the unlabeled images respectively, and $\lambda$ represent the weight of unsupervised loss in the total loss function.\par
At the beginning of training, the student model and teacher model adopt random initialization. For labeled data, the student network uses its ground truth to calculate the loss $L_{\text {sup }}$ as (\ref{math2}) and update the student model parameters by the gradient descent method. For unlabeled data, the teacher network first deduces the result of weakly augmented unlabeled data, then filtering out the low-quality results according to the confidence threshold $\sigma$. The retained high-quality predictions is considered to be the pseudo labels for unlabeled data. Then, the student calculates the unsupervised loss $L_{\text {unsup }}$ of unlabeled data as (\ref{math3}), and updates the parameters by the gradient descent method. Finally, the teacher network is updated with the exponential moving average (EMA) for the student model.\par
\begin{equation}\label{math2}
L_{\text {sup }}=\frac{1}{N_{\text {label }}} \sum_{t=1}^{N_{\text {label }}} L_{c l s}\left(v_{t}, \hat{v}_{t}\right)+\sum_{t=1}^{N_{\text {label }}}L_{r e g}\left(\theta_{t}, \hat{\theta}_{t}\right)
\end{equation}
\begin{equation}\label{math3}
L_{\text {unsup }}=\frac{1}{N_{\text {unlabel }}} \sum_{t=1}^{N_{\text {unlabel }}} L_{c l s}\left(v_{t}, v_{t}^{*}\right)+\sum_{t=1}^{N_{\text {unlabel }}} L_{r e g}\left(\theta_{t}, \theta_{t}^{*}\right)
\end{equation}
where $N_{\text {label }}$ indicates the number of positive samples in the labeled data, $N_{\text {unlabel }}$ indicates the number of positive samples in pseudo data with retained classification scores above the threshold $\sigma$. $v_{t}$ and $\hat{v}_{t}$ represent the student predicted category and ground truth of the t-th positive sample, $\theta_{t}$ and $\hat{\theta}_{t}$ represent the student predicted regression result and corresponding target of the t-th positive sample. $v_{t}^{*}$ and $\theta_{t}^{*}$ represent the category and box of the positive sample whose classification score from pesudo labels is higher than the threshold $\sigma$.
\subsection{Adaptive difficult sample oriented method (ADSO)}
The quality of pseudo labels determines the performance of SSL model, in our framework, pseudo labels with low classification scores will be filtered out with confidence threshold $\sigma$ to ensure the quality of pseudo labels. Neither the high confidence threshold setting nor low confidence threshold setting is suitable for the semi-supervised framework based on single-stage detectors. For the SSOD method based on single-stage detector, directly using the same setting as the two-stage detector (confidence threshold = 0.7 in \cite{2021Unbiased}) will greatly increase the number of negative samples, and only few part can be predicted in positive samples\cite{2022S4OD}. This imbalance will result in not getting enough pseudo labels, which limits the performance of the semi-supervised single-stage detector. On the contrary, low threshold threshold will easily bring in more low-quality pseudo labels with noise. When we adopt confidence $\sigma$ = 0.5, as shown in Figure \ref{figure3}, AP50 decreases to a great extent with the increase of training iterations. We believe with training iterations increasing, Teacher have been able to generate enough pseudo labels with confidence greater than $\sigma$. The number of pseudo labels has been much more than that in the early stage of training, but limited to low threshold settings, the classification score of pseudo labels has not been greatly increased with the increase of training iterations. We need to adopt a more strict confidence screening strategy in the middle stage of training and provide higher score pseudo labels.\par
We propose the adaptive difficult sample oriented (ADSO) moudle, which inherits the advantages of the flexibility of the end-to-end framework and fully utilizes the information provided by the teacher model. Specifically, we traverse the pseudo labels after confidence filtering to evaluate the reliability of each label generated by teacher. This labels generated by teacher are regarded as real pseudo labels. In this process, deal with the confidence of each pseudo box without considering background boxes. The processing function is shown in the Figure \ref{figure7}:
\begin{figure}
	\centering 
	\includegraphics[scale=0.2]{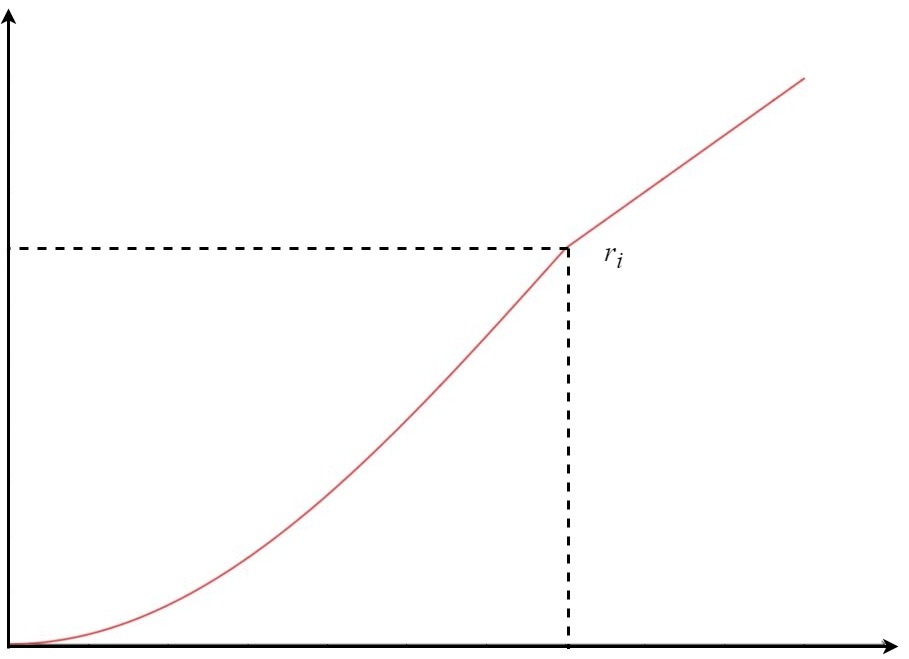}
	\caption{Deal with the confidence score predicted by teacher. The confidence scores predicted by teacher are processed to reduce the confidence of input pseudo labels with inferior confidence.}
	\label{figure7}
\end{figure}
when the confidence is close to the confidence threshold setting, the confidence is reduced to a lower value through the function, at this time when the confidence is already high, we take the current confidence as the evaluation standard of prospect reliability in (\ref{math4}), so that the classification loss function of unsupervised part is as (\ref{math5}). Liu \textit{et al.}\cite{2021Unbiased} have proved by a large number of experiments that selecting 0.7 as the confidence threshold in the two-stage network will achieve better results. Therefore, we select $r_ {i} $=0.7 to make the pseudo boxes' classification score approach the two-stage network. When the confidence is low, a penalty item will be added to the loss calculation of the pseudo box which is similar to the hard negative sample mining method. We find that this method makes the classification score after a long iteration higher than without ADSO as shown in Figure \ref{figure4}.\par
\begin{equation}\label{math4}
\omega_{i}=y(x)= \begin{cases}r_{i} \sin \left(\frac{\pi}{2}  \frac{x-r_{i}}{r_{i}  }\right)+r_{i} & : x<r_{i} \\ x & : r_{i}<x<1\end{cases}
\end{equation}
\begin{equation}\label{math5}
L_{\mathrm{unsup}}^{c l s}=\frac{1}{N_{\text {unlabel }}^{f g} \omega_{i}} \sum_{i=1}^{N_{i}^{f g}} L_{c l s}\left(v_{t}, v_{t}^{*}\right)
\end{equation}
where $L_{\mathrm{unsup}}^{c l s}$ is unsupervised classification loss, $x$ indicates the classification score of the i-th box, $\omega_{i}$ indicates the classification score after function transformation as (\ref{math4}).\par
\begin{figure}
	\centering 
	\includegraphics[scale=0.4]{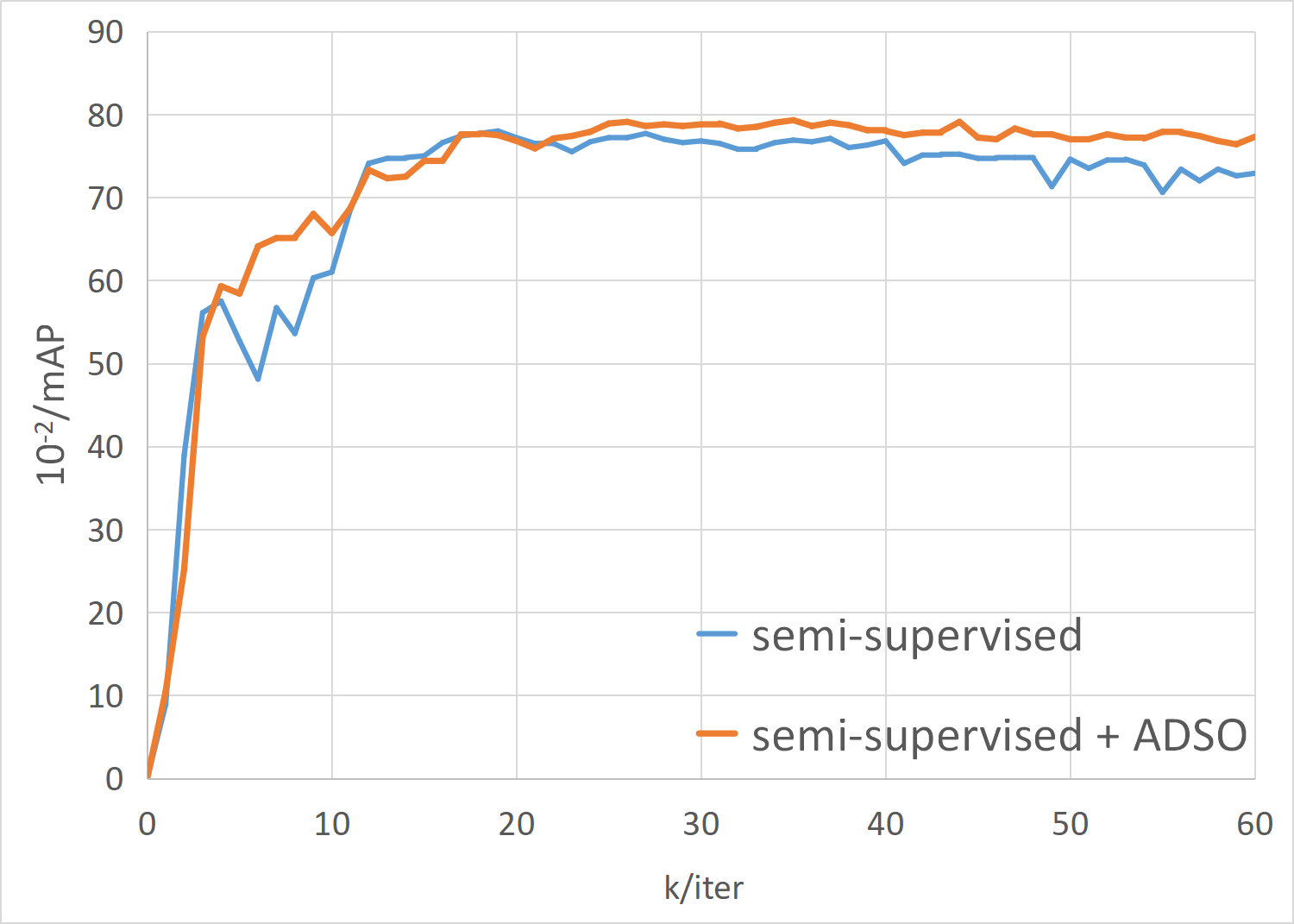}
	\caption{Changes of AP50 with training time before and after adding  ADSO module. ADSO makes the model's AP50 more stable.}
	\label{figure3}
\end{figure}
\subsection{Fusion Box}
Different from the classification quality evaluation of pseudo labels, the regression quality of pseudo boxes is a difficult index to evaluate. We visualized the pseudo labeled images of the teacher model in the training process, As shown in Figure \ref{figure9}. Because the prediction accuracy of the teacher model is not high enough, multiple pseudo boxes will be predicted around ground truth boxes of the unlabeled images. Most of these pseudo boxes can not provide reliable and accurate positioning information for student but after merging these pseudo boxes, the new box location is closer to the ground truth box. Therefore, we introduce Fusion Box module to reduce the inaccurate location impact of pseudo boxes. Specifically, for an unlabeled image predicted by teacher model, after confidence filtering, Fusion Box module select whether to synthesize each other through the similarity $\xi$ and the pseudo algorithm of Fusion Box is as \textbf{Algorithm 1}.\par
\begin{figure}
	\centering 
	\includegraphics[scale=0.8]{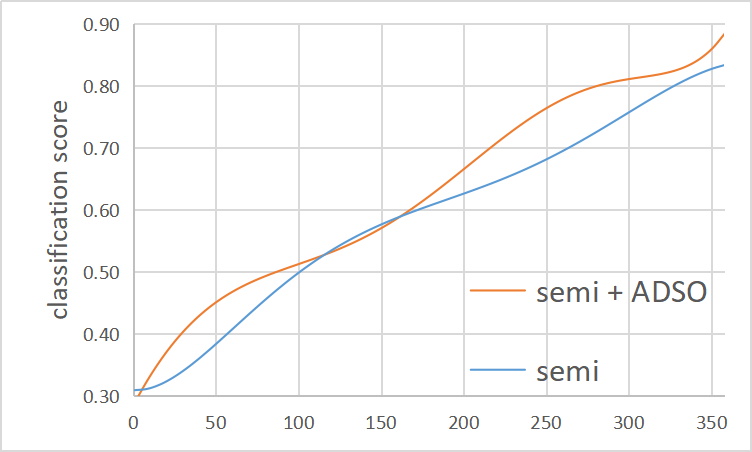}
	\caption{The labels confidence of prediction boxes of 358 Val images predicted by semi supervised learning framework with and without ADSO is to gather statistics. The method of adding ADSO module can predict boxes with higher confidence}
	\label{figure4}
\end{figure}
Fusion Box collects the euclidean distance of the center point between each other for all the pseudo boxes generated in each iteration, and then calculates the similarity between each other. The two boxes whose similarity is less than the threshold $\mu$ are combined into a new pseudo box. The calculation process of the similarity $\xi$ of these pseudo boxes can be expressed by (\ref{math6}).

%	%% 写算法伪代码或者流程的前期准备
%% 写算法伪代码或者流程的前期准备
\renewcommand{\algorithmicrequire}{\textbf{Input:}}  % Use Input in the format of Algorithm
\renewcommand{\algorithmicensure}{\textbf{Output:}} % Use Output in the format of Algorithm

\begin{algorithm}\label{algorithm}
	\caption{Fusion Box } % 名称
	\begin{algorithmic}[1]
		\Require	
		Prediction $  (B o x_{(x, y)}, Score)$, confidence threshold $\sigma$, Fusion Box threshold $\mu$
		\Ensure
		Fusion prediction
		\State {Box$_{(x, y)} \leftarrow$ Box$_{(x, y)}[$ Score$\geq \sigma]$, Score$\leftarrow$Score$[$Score $\geq \sigma]$}
		\If  {$\operatorname{Size}\left(\operatorname{Box}_{(x, y)}\right)>1$ } 
		\For  {$n=0$ to $\operatorname{Size}\left(\operatorname{Box}_{(\mathrm{x}, \mathrm{y})}\right)$}\State$\text{Centre}[n]=\operatorname{mean}\left[\operatorname{Box}_{(x, y)}[n]\right]$
		\EndFor
		\For {$m=0$ to $\operatorname{Size}\left(\operatorname{Box}_{(\mathrm{x}, \mathrm{y})}\right)$}\For{$n=0$ to $\operatorname{Size}($Center$)$}
		\State	$\text {$\xi$} =\operatorname{Center}[m]-H[n]$
			\If {$
						\text { $\xi$ }< \mu
						$} \State $Box_{(x, y)}[m]=Bo x_{(x, y)}[n] \cup B o x_{(x, y)}[m]
				   $\State$
			     	 Del\ Bbox_{(x,y)}[n]
				   $\State$
			    	Del\ Center[n]
			     	$		
			   	\EndIf
				    \EndFor	
			    	\EndFor
		\EndIf	
	\end{algorithmic}
\end{algorithm}
\begin{figure}
	\centering 
	\includegraphics[scale=0.25]{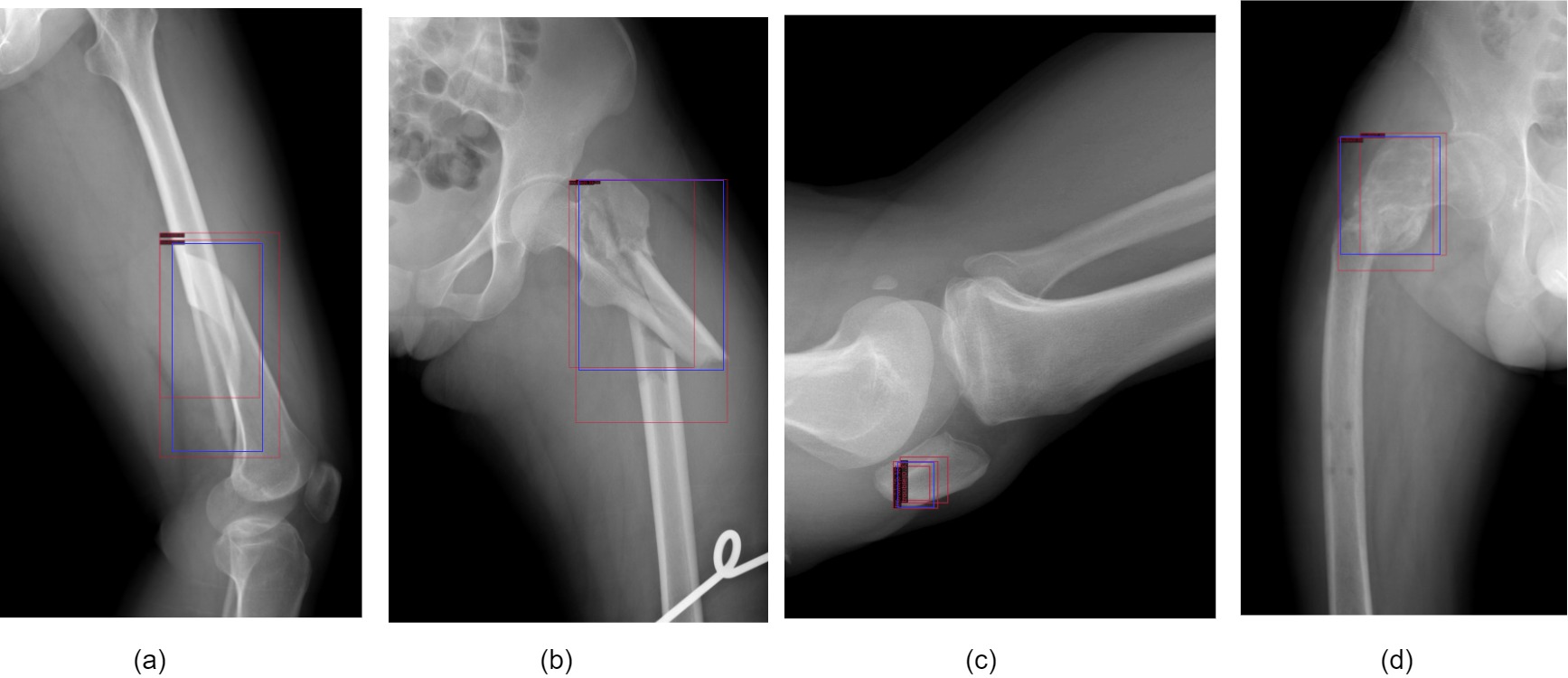}
	\caption{blue: annotation; red: pseudo label.}
	\label{figure9}
\end{figure}
\begin{equation}\label{math6}
\text { $\xi$ }=\frac{|\operatorname{mean}(\mathrm{Bbox}[\mathrm{n}])-\operatorname{mean}(\mathrm{Bbox}[\mathrm{m}])|^{2}}{\operatorname{scale}(w)+\operatorname{scale}(h)}
\end{equation}
where$
\text { mean(Bbox }[\mathrm{n}])
$ indicates that the center coordinates of each pseudo box. $\operatorname{scale}(w)$ and $\operatorname{scale}(h)
$ represent the width and height of the whole image scaled after data augmentation. After Fusion Box module, the unsupervised regression loss function is calculated as follows:
\begin{equation}\label{math7}
L_{\mathrm{unsup}}^{reg}
=\frac{1}{N_{p o s_{\text {fusion }}}}\sum_{t=1}^{N_{p o s_{\text {fusion }}}}L_{r e g}\left(\theta_{t},\theta_{t}^{*}\right)
\end{equation}
where $L_{\mathrm{unsup}}^{reg}$ is unsupervised regression loss, $N_{p o s_{\text {fusion }}}$ indicates the number of pseudo boxes after confidence filtering and Fusion Box, $\theta_{t}^{*}$ 
represents the regression target after Fusion Box.

\subsection{Dex encoder}
As shown in Figure \ref{figure1}, we design the Dex encoder  to replace the dilated encoder in YOLOF.
Dilated encoder stacks standard convolution and dilated convolution, then merge the original feature to the feature map containing expanded receptive field. For reducing the parameter quantity, we cancel the dilated block with dilation rate = 2 in Dex encoder and choose to add 1×1 convolution after Deformable convolution v2 (DCNv2) module\cite{Zhu_2019_CVPR}, the convolution layer reduces channels to 128 and maintains the reduced number of channels throughout the network.\par
\begin{figure}
	\centering 
	\includegraphics[scale=0.2]{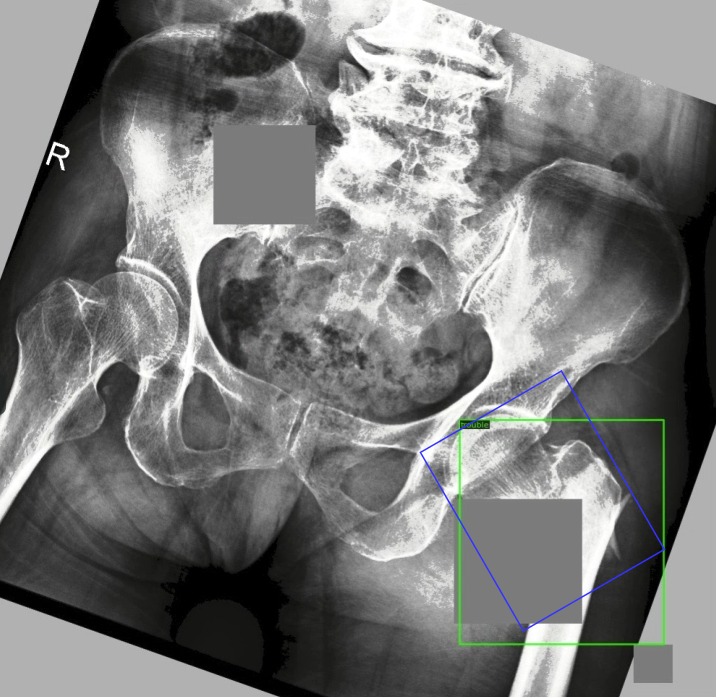}
	\caption{The augmented image after cutout and rotation.The pseudo labels (green) labeled by the teacher on the augmented image will not only have position deviation but also have angle difference with ground truth box (blue)}
	\label{figure6}
\end{figure}
\begin{figure}
	\centering 
	\includegraphics[scale=0.4]{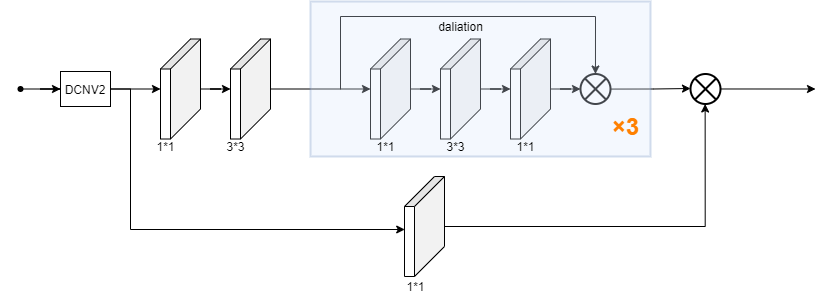}
	\caption{structure of Dex encoder }
	\label{figure1}
\end{figure}

Data augmentation need to be used for the input images in SSOD. The unlabeled images after different data augmentation will be collected to mark pseudo labels in the unsupervised section. Inputting these deformed images to network will lead to coordinates and angle changes in the ground truth box corresponding to Figure \ref{figure6} while the deformed box will seriously affect the network training. Compared with the traditional fixed window convolution, deformable convolution \cite{dai2017deformable} can effectively deal with gtbox deformation including box movement, size scaling and rotation, because its local receptive field is learnable and oriented to the whole image. Deformable convolution is to increase the spatial sampling position of additional offset and does not need additional supervision. Therefore, we choose to apply
DCNv2 to YOLOF.\par

\begin{figure}
	\centering 
	\includegraphics[scale=0.4]{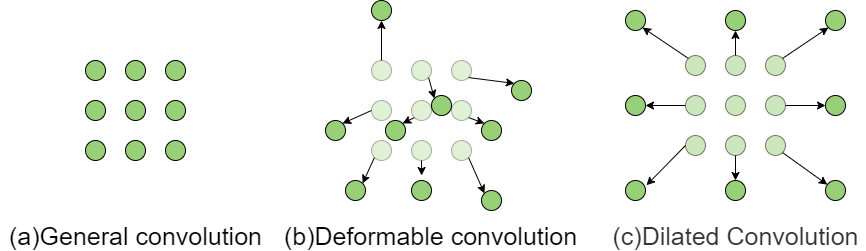}
	\caption{The structure of different convolution kernel}
	\label{figure10}
\end{figure}
As shown in Figure \ref{figure10}, (a) is fixed window convolution, (b) is deformation convolution, and saturation color points is the actual sampling position of convolution kernel, which is offset from the standard position. (c) is dilated convolution which can be regarded as the special form of deformation convolution, has
the ability to expand the receptive field. The convolution kernel used to generate the output
feature and the deformation convolution kernel used to generate the offset are synchronously learned in
DCNv2 while the offset is obtained by back propagation using interpolation algorithm.

\section{Experiments}
\subsection{Dataset and Evaluation protocol}
We validate our method on the thighbone fracture dataset\cite{2019Thigh} which was collected from the radiology department of Linyi people's hospital. All X-ray images are produced by the latest digital radiography (DR) technology. The dataset consists of 3842 thigh fracture images in 24 bit JPG format. The train set contains 3484 labeled images. In addition, The test set of 358 images is provided for serious performance verification. The following settings are adopted for performance verification:\par
Referring to the verification method of STAC\cite{sohn2020simple}, we use 1\%, 5\% and 10\% of trainset as labeled training data, and unselected images in trainset  as unlabeled data. In addition, our knowledge distillation result is compared with the supervised thighbone fracture detection methods proposed by Guan \textit{et al.}\cite{2019Thigh} and Wang \textit{et al.}\cite{wang2021parallelnet} on this dataset to prove our advantage, following the convention to report the performance of val with mAP and AP50 as evaluation indicators.
\subsection{Implementation Details}
The experiment was conducted on 4 NVIDIA geforce RTX 1080ti. We use Resnet-50 as feature extraction network to compare with previous methods. The backbone is initialized on ImageNet with pre-trained weight. Our implementation and hyperparameters are based on mmdetection\cite{chen2019mmdetection}, anchors with 5 scales and 1 aspect ratios are used. In the comparison between using partial labeled data sets and being compared with supervised algorithms, the training parameters under the two settings are slightly different because there is a large discrepancy in the amount of labeled training data.\par
\textbf{Partial annotation data.} the model is trained on 4 GPUs for 60K iterations, and each GPU has 8 images. During SGD training, the initial learning rate is 0.02, and the learning rate of backbone part is set to 0.005, which is divided by 10 in the 1k-th iteration and 1.5k-th iteration. The warmup phase is delayed to 1500iter. The weight attenuation and momentum are set to 0.0001 and 0.9, respectively. The confidence threshold is set to 0.5, the data sampling ratio SR is set to 0.25, and gradually decreases to 0 in the last 1K iterations.\par
\textbf{Compared with supervised algorithms.} the data sampling ratio SR is set to 1, other settings are the same as the experience of partial annotation data.\par
In order to estimate the reliability of Fusion Box, the Fusion Box threshold $\mu$ is selected as 0.05 according to the size of the input image and select the fused pseudo box for box regression. In addition, we use weak data augmentation for teacher model, strong data augmentation for student model as shown in Table \ref{1}. (Unlabeled (T) and Unlabeled (S) respectively represent the data augmentation used for the unlabeled images input to the teacher and student models while Labeled represents the data augmentation used for labeled images. Weak Aug and Strong Aug respectively represent the weak data augmentation and strong data augmentation.)\par
\begin{table}
	\centering
	\caption{Data augmentation implementation details for our semi-supervised approach.}\label{1} 
	\begin{tabular}{@{}llll@{}}
		\toprule
		& \textbf{Labeled}              & \textbf{Unlabeled (S)}         & \textbf{Unlabeled (T)} \\ \midrule
		\textbf{Weak Aug}                                     &                               &                               &                         \\
		Scale jitter                                                   & short edge  $\in$  (0.5, 1.5)       & short edge $\in$ (0.5, 1.5)       & short edge $\in$ (0.5, 1.5) \\
		Horizontal Flip                                                & p=0.5                         & p=0.5                         & p=0.5                   \\
		\textbf{Strong Aug}                                   &                               &                               &                         \\
		Contrast jitter                                                & p=0.2, ratio $\in$ (0, 1)         & p=0.2, ratio $\in$ (0, 1)         & p=0.25, ratio  $\in$ (0, 1)  \\
		Solarize jitter & p=0.1,, ratio  $\in$ (0, 1)        & p=0.1,, ratio  $\in$ (0, 1)        & -                       \\
		Color Jitter                                                   & p=0.1,ratio=(0.4,0.4,0.4,0.1) & P=0.1,ratio=(0.4,0.4,0.4,0.1) & -                       \\
		Brightness jitter                                              & p=0.1, ratio  $\in$  (0, 1)         & p=0.1, ratio  $\in$  (0, 1)         & -                       \\
		Sharpness jitter                                               & p=0.1, ratio  $\in$  (0, 1)         & p=0.1, ratio  $\in$ (0, 1)         & p=0.25, ratio  $\in$  (0, 1)  \\
		Posterize                                                      & p=0.1                         & p=0.1                         & -                       \\
		Equazlize                                                      & p=0.1                         & p=0.1                         & p=0.25                  \\
		Rotate                                                         & -                             & p=0.3, angle  $\in$  (0, 30◦)       & p=0.3, angle  $\in$ (0, 30◦) \\
		SHIFT                                                          & -                             & p=0.3, angle  $\in$ (0, 30◦)       & p=0.3, angle  $\in$  (0, 30◦) \\
		Cutout                                                         & -                             & ratio  $\in$  (0.05, 0.2)           &                         \\ \bottomrule
	\end{tabular}\par
\end{table}
\subsection{System Comparison}
\begin{figure}
	\centering 
	\includegraphics[scale=0.5]{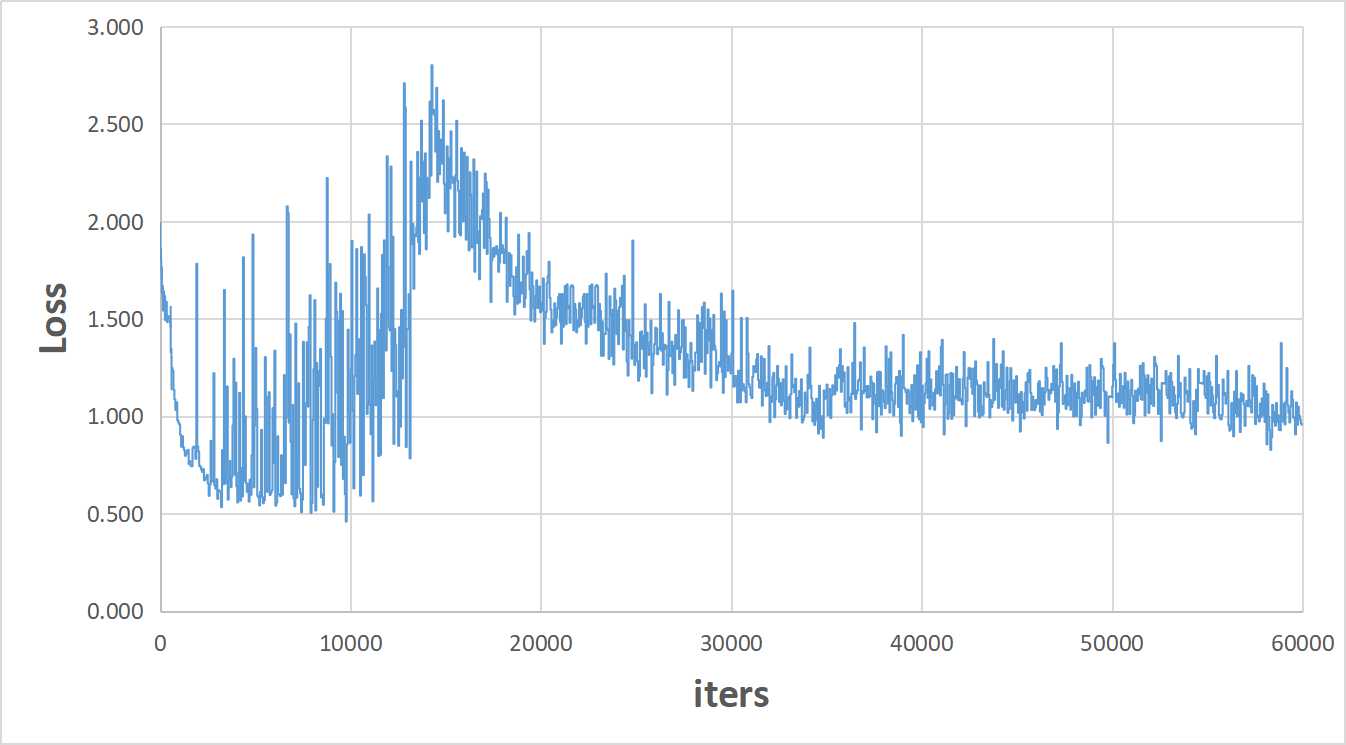}
	\caption{Train loss. In the early stage of training, the network accuracy is insufficient and it is difficult to obtain pseudo labels. Most of the semi-supervised networks are trained with labeled data, and the loss gradually decreases. With the increase of network training times, the teacher model labels enough pseudo labels, and the student model carries out more unsupervised training, then the loss gradually increases. In the later stage of training, with the quality of pseudo labels getting higher, the semi-supervised learning of the system reaches saturation.}
	\label{figure11}
\end{figure}

\begin{figure}
	\centering 
	\includegraphics[scale=0.05]{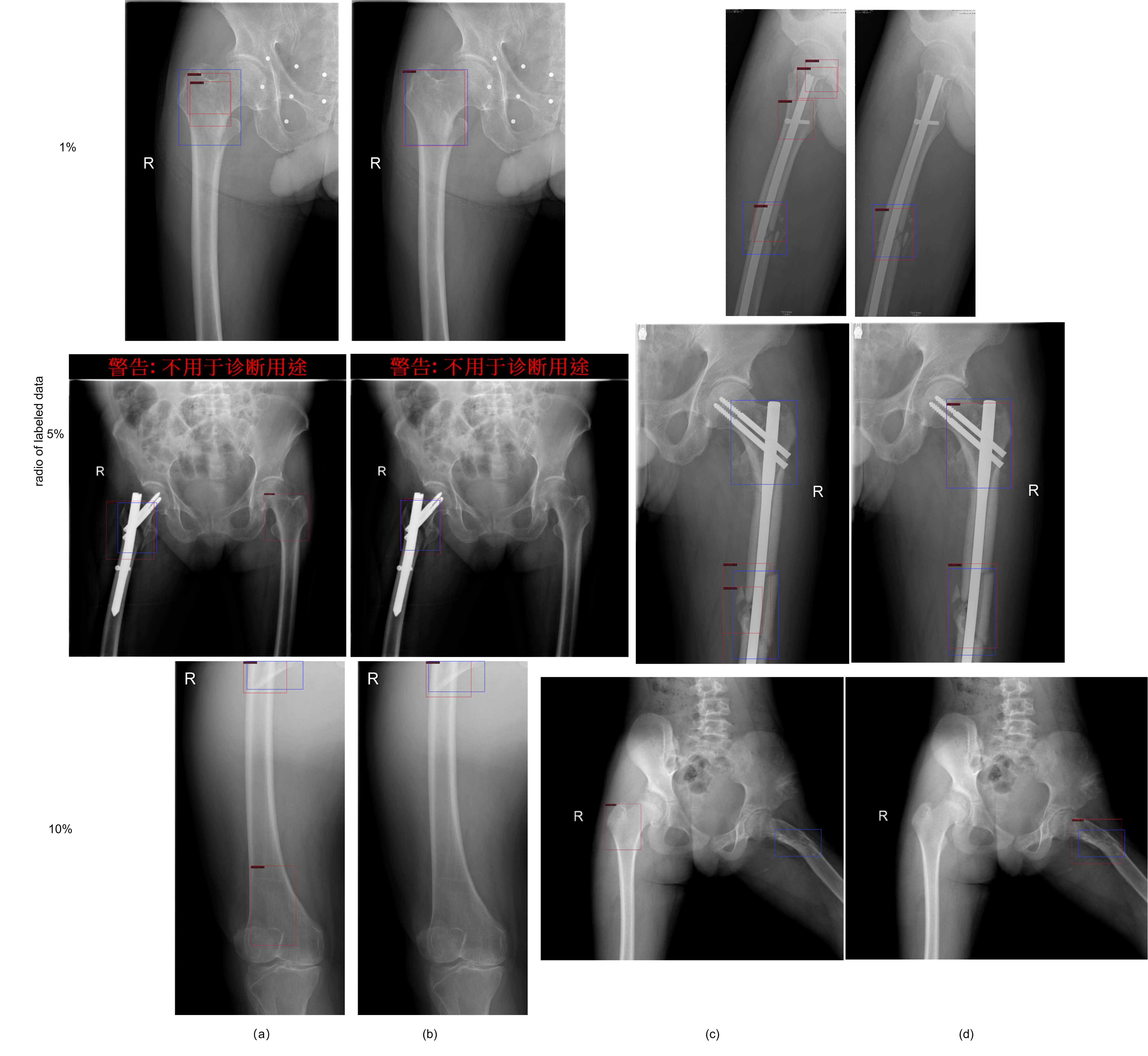}
	\caption{(a), (c) represent YOLOF baseline forecast results.(b), (d) represent prediction results of semi-supervised method (blue: annotation; red: detection results).}
	\label{figure5}
\end{figure}

In this section, we compare our method with other state-of-the-art methods proposed in recent years. We first evaluate on Partial annotation data setting and compare our method with the results in STAC, Soft teacher and Unbiased teacher. Also the baseline of YOLOF and FasterRcnn is compared with our semi-supervised method in Table \ref{2}. We found that our method performed better than other methods in fracture detection. Specifically, the mAP of our method is 15.7\%, 22.5\% and 11.0\% higher than the baseline on the 1\%, 5\% and 10\% data sets, at the same time 2.4\%, 0.2\% and 2.3\% higher than the previous methods. We evaluate the loss of model training and show the
results in Figure \ref{figure11}. In order to compare the prediction results more intuitively, we visualized the baseline prediction using YOLOF and the prediction results of our semi-supervised framework in Figure \ref{figure5}.\par
Then we compared our performance of SSL framework on full dataset and 50\% labeled dataset in Table \ref{9} and Table \ref{10}. We only use 50\% data for semi-supervised training, and the AP exceeds Cascade r-cnn, a two-stage network which parameters are much  more than YOLOF. By using Dex encoder, we only need 3 divided blocks to increase the mAP of YOLOF by 1.6\% and AP75 by 2.6\%, which proves the effectiveness of Dex encoder. In addition, using our semi-supervised framework to distill the knowledge of the single-stage network can further improve 2.6\% AP50 and 9.6\% AP75. Compared with the algorithm for fracture detection and the current popular object detection network, our semi-supervised framework has higher accuracy. In addition, by introducing our semi-supervised framework, our model achieves the same performance as the fully supervised training of two-stage detectors with simpler network structure and fewer parameters.
We also compare the model inference speed on test set and moel complexity with the latest two methods in Table \ref{4}.

\begin{table}[]
			\caption{The results compared with baseline and other semi-supervised methods in 1\%, 5\% and 10\% labeled data.}\label{2} 
	\centering
	
	\begin{tabular}{@{}lllllllll@{}}
		\toprule
		Method                                                                     & \multicolumn{2}{l}{Backbone}                   & \multicolumn{2}{c}{1\%}        & \multicolumn{2}{c}{5\%}        & \multicolumn{2}{l}{10\%}       \\ \midrule
			& \multicolumn{2}{l}{ResNet-50} & mAP           & AP50         & mAP           & AP50         & mAP           & AP50         \\
			\begin{tabular}[c]{@{}l@{}}Supervised baseline (FasterRcnn)\end{tabular} & \multicolumn{2}{l}{}                           & 6.5           & 22.3          & 12.1          & 38.8          & 28.4          & 68.3          \\
			\begin{tabular}[c]{@{}l@{}}Supervised baseline (YOLOF)\end{tabular}      & \multicolumn{2}{l}{}                           & 6.1           & 22.6          & 13.8          & 39.1          & 29.6          & 69.6          \\
			STAC                                                                       & \multicolumn{2}{l}{}                           & 11.6          & 31.2          & 19.7          & 55.4          & 33.8          & 72.9          \\
			Softteacher                                                                & \multicolumn{2}{l}{}                           & 19.2          & 45.6          & 36.1          & 71.6          & 38.3          & 75.1          \\
			Unbiased teacher                                                           & \multicolumn{2}{l}{}                           & 19.8          & 53.4          & 31.7          & 72.7          & 37.7          & 78.1          \\
			\textbf{Ours}                                                              & \multicolumn{2}{l}{}                           & \textbf{22.2} & \textbf{53.9} & \textbf{36.3} & \textbf{75.0} & \textbf{40.6} & \textbf{80.1} \\ \bottomrule
		\end{tabular}

\end{table}
% \usepackage{booktabs}
% \usepackage{booktabs}
\begin{table}[]
	\centering
	    \caption{Performance of semi-supervised framework on full labeled datasets.}\label{9}
	\begin{tabular}{@{}llll@{}}
		\toprule
		Algorithm                                 & Backbone network        & AP50          & AP75                \\ \midrule
		\multicolumn{4}{c}{\textbf{Current object detection algorithm}}                                 \\
		YOLOF                                     & Resnet-50               & 82.2          & 39.6                \\
		FCOS                  & Resnet-101              & 85.4          & 42.7                \\
		Empirical attention   & Resnet-50               & 86.1          & 43.2                \\
		GHM        & Resnet-50               & 86.5          & 43.8                \\
		GCNet                & Resnet-50               & 85.4          & 44.6                \\
		FPN                   & Resnet-101              & 86.3          & 48.8                \\
		Cascade R-CNN & Resnet-50               & 85.0          & 49.7                \\
		\multicolumn{4}{c}{\textbf{Object detection algorithms in the thighbone fracture dataset}}                  \\
		YOLOF(Dex encoder+3 dilated blocks)       & Resnet-50               & 83.8(+1.6)    & 42.2(+2.6)          \\
		framework(M et al,2019)                   & network(M et al,2019\cite{wang2021parallelnet})   & 87.3          & 47.6                \\
		framework(Bin et al,2022)                 & network(Bin et al,2022\cite{0Automatic}) & \textbf{88.9} & 52.6                \\
		\textbf{Our semi-supervised method}      & Resnet-50               & 86.2(+2.6)    & \textbf{52.6(+9.6)} \\ \bottomrule
	\end{tabular}
\end{table}
% \usepackage{booktabs}
\begin{table}[]
		\caption{Performance of semi-supervised framework on full labeled dataset and 50\% labeled datatset.}\label{10}
	\centering
	\begin{tabular}{@{}lll@{}}
		\toprule
		Algorithm                                                   & AP50 & AP75 \\ \midrule
		\multicolumn{1}{c}{ our semi-supervised framework(50\% data)} & 85.0  & 51.1  \\
		\multicolumn{1}{c}{our semi-supervised framework(full data) }                   & 86.2  & 52.6  \\ \bottomrule
		\end{tabular}
	\end{table}
\begin{table}[]
	\centering
		\caption{Ablation results with different modules.}\label{3} 
	\begin{tabular}{@{}lllll@{}}
		\toprule
		ADSO & Fusion Box & Dex encoder & mAP           & AP50         \\ \midrule
		&            &             & 37.3          & 77.7          \\
		\checkmark    & \textbf{}  & \textbf{}   & 39.0          & 78.1          \\
		& \checkmark          &             & 38.4          & 78.0          \\
		&            & \checkmark           & 38.4          & 78.6          \\
		\checkmark    &            & \checkmark           & 39.9          & 79.2          \\
		\checkmark    & \checkmark          & \checkmark           & \textbf{40.6} & \textbf{80.1} \\ \bottomrule
	\end{tabular}

\end{table}\par
\begin{table}[]
		\caption{Ablation results of  dilated block quantity. The number in () indicates each dilated block’s dilated rate.}\label{5}
	\centering
	\begin{tabular}{@{}lll@{}}
		\toprule
		method                                         & mAP           & AP50         \\ \midrule
		YOLOF(2,4,6,8)                                 & 29.4          & 69.3          \\
		Dex encoder+1 dilated block(4)                 & 29.9          & 69.9          \\
		Dex encoder+2 dilated blocks(4,6)              & 30.0          & 70.6          \\
		Dex encoder+3 dilated blocks(4,6,8)            & 30.0          & 71.9          \\
		\textbf{Dex encoder+4 dilated blocks(2,4,6,8)} & \textbf{31.4} & \textbf{72.5} \\
		Dex encoder+5 dilated blocks(2,4,6,7,8)        & 30.8          & 71.3          \\ \bottomrule
	\end{tabular}

\end{table}\par
\begin{table}[]
		\caption{Computational complexity and inference speed of different models in 358 val images.}\label{4}
	\centering
	\begin{tabular}{@{}lll@{}}
		\toprule
		Method        & FLOPs            & FPS           \\ \midrule
		Unbiased teacher & 204.13G          & 22.4          \\
		Soft teacher      & 202.31G          & 16.0          \\
		\textbf{Ours}    & \textbf{101.57G} & \textbf{27.6} \\ \bottomrule
	\end{tabular}
 
\end{table}

% \usepackage{booktabs}
\begin{table}[]
	\centering
		\caption{Ablation results of different confidence thresholds.}\label{21}
	\begin{tabular}{@{}lll@{}}
		\toprule
		Threshold $\sigma$    & mAP           & AP50         \\ \midrule
		0.4          & 39.8          & 79.3          \\
		\textbf{0.5} & \textbf{40.6} & \textbf{80.1} \\
		0.55         & 40.1          & 79.7          \\
		0.6          & 39.6          & 79.1          \\ \bottomrule
	\end{tabular}

\end{table}
% \usepackage{booktabs}
\begin{table}[]
	\caption{Ablation results of Fusion Box thresholds.}\label{22}
	\centering
	\begin{tabular}{@{}lll@{}}
		\toprule
		Threshold $\xi$ & mAP           & AP50         \\ \midrule
		0.04            & 39.7          & 79.6          \\
		0.05            & \textbf{40.6} & \textbf{80.1} \\
		0.06            & 39.8          & 79.8          \\ \bottomrule
	\end{tabular}
\end{table}

\subsection{Ablation Studies}
In this section, we validate key designs. All ablation experiments are performed on YOLOF using 10\% labeled dataset.\par
\textbf{The influence of critical design in semi-supervised model.} Table \ref{3} shows the impact of different modules on the performance of the model. When all three modules are adopted, the model performance can reach the best. Compared with the original SSL model, mAP is increased by 3.3\% and AP50 by 2.4\%. The validity of modules is proved.\par
\textbf{The impact of confidence filter threshold.}
We select different confidence thresholds to compare the performance of our semi-supervised model in Table \ref{21}. When we select a threshold of 0.5, the performance of the model reaches the best. When we select a threshold of 0.5 ± 0.1, a high threshold will lead to a faster decline in the mAP, indicating that our semi-supervised model is sensitive to a high threshold. However, whether we select a high threshold or a low threshold, the accuracy of the model does not decline much, indicating that ADSO module has a certain adjustment function to the confidence threshold.\par
\textbf{The impact of dilated block quantity.}We compared the effect of  different number of expansion blocks in Table \ref{5}. Considering the balance between accuracy and parameter quantity, we finally choose to use 3 dilated blocks on the Dex   encoder.\par
\textbf{The impact of Fusion Box threshold.} In Table \ref{22}, we study the box regression Fusion Box threshold. The best performance is achieved when the threshold is set to 0.05.\par

\section{Conclusions}
In this paper, a semi-supervised object detection  method based on the single-stage network is proposed to train neural networks with limited labeled data and a large number of unlabeled data. In the end-to-end training, we propose three modules: ADSO, Fusion Box and Dex encoder. We improve the object detection network to promote the effective use of teacher model. For the detection of thighbone fracture in clinical application, the model has high accuracy. A large number of experiments on the thighbone fracture dataset show that semi-supervised method has broad application prospects in the field of medical images. We hope that our work can help surgeons improve the efficiency of diagnosing diseases. \par

\section{Reference}
\bibliography{mybib}

\end{document}